\documentclass[conference]{IEEEtran}
\usepackage[utf8]{inputenc}
\usepackage{cite}
\usepackage[pdftex]{graphicx}
\usepackage[cmex10]{amsmath}
\usepackage{algorithmic}
\usepackage{array}
\usepackage[tight,footnotesize]{subfigure}
\usepackage{multirow}
\usepackage{stfloats}
\usepackage{xcolor}
\usepackage[font=small,skip=2pt]{caption}
\usepackage{multicol}
\usepackage{siunitx}
\usepackage{footnote}
\usepackage{setspace}

\title{Delay lines test method for the Blade Template}
\author{\IEEEauthorblockN{Felipe A. Kuentzer\IEEEauthorrefmark{1}\IEEEauthorrefmark{2}, Leonardo R. Juracy\IEEEauthorrefmark{3}, Matheus T. Moreira\IEEEauthorrefmark{4} and Alexandre M. Amory\IEEEauthorrefmark{3}}
\IEEEauthorblockA{\IEEEauthorrefmark{1}IHP – Leibniz-Institut für innovative Mikroelektronik \\
\IEEEauthorrefmark{2}Department of Computer Science, University of Potsdam \\
\IEEEauthorrefmark{3}Faculty of Informatics, PUCRS University\\
\IEEEauthorrefmark{4}Chronos Tech, Research and Development\\
kuentzer@ihp-microelectronics.com, leonardo.juracy@acad.pucrs.br, matheus@chronostech.com, alexandre.amory@pucrs.br}}

\begin{document}

\setstretch{0.96}

\maketitle

\begin{abstract}
This paper proposes a method for testing the delay lines of Blade asynchronous timing resilience template. It consists of an offline test method, with low area overhead, for measuring the internal propagation delay of the delay lines with external testers. 
\end{abstract}

\vspace{-0.8em}

\section{Introduction}

Asynchronous bundled-data designs, in general, must incorporate delay lines in their control signals to compensate for data path propagation through computational logic. The correctness of delay lines is crucial for the system to work correctly. However, the delay lines are subject to variability and faults. 

In bundled-data designs, the circuit is faulty if the data path becomes slower or if the delay lines become faster than the timing specifications. In the literature, there are not many works that address the testability of delay lines in bundled-data designs. The work of Sato \cite{SATO15} relies on scannable data path and two-pattern delay test methodology to measure the delay of the combinational block, thus the method does not take delay measurements of the delay lines itself. Timing information is captured by a time to digital converter and used later to configure the delay lines to a value that is long enough to compensate for the data path delay. Moreover, applying Sato's \cite{SATO15} approach would incur in significant area overhead due to the need of a scannable data path.

Blade \cite{BLADE15} is an asynchronous bundled-data template with timing resilience capabilities, that can recover from timing violation in the data path. Blade relies on the delay lines as a timing reference to detect the timing violations. So, this paper proposes a method for testing Blade's delay lines. It consists of an offline test method for measuring the propagation delay of the delay lines of the manufactured circuit. The results show that the method has low area overhead and a simple DfT circuitry. On the other hand, it requires an accurate external tester to precisely measure the delay of the delay lines via external pins.

\begin{figure*}[htb]
    \centering \includegraphics[scale=0.27]{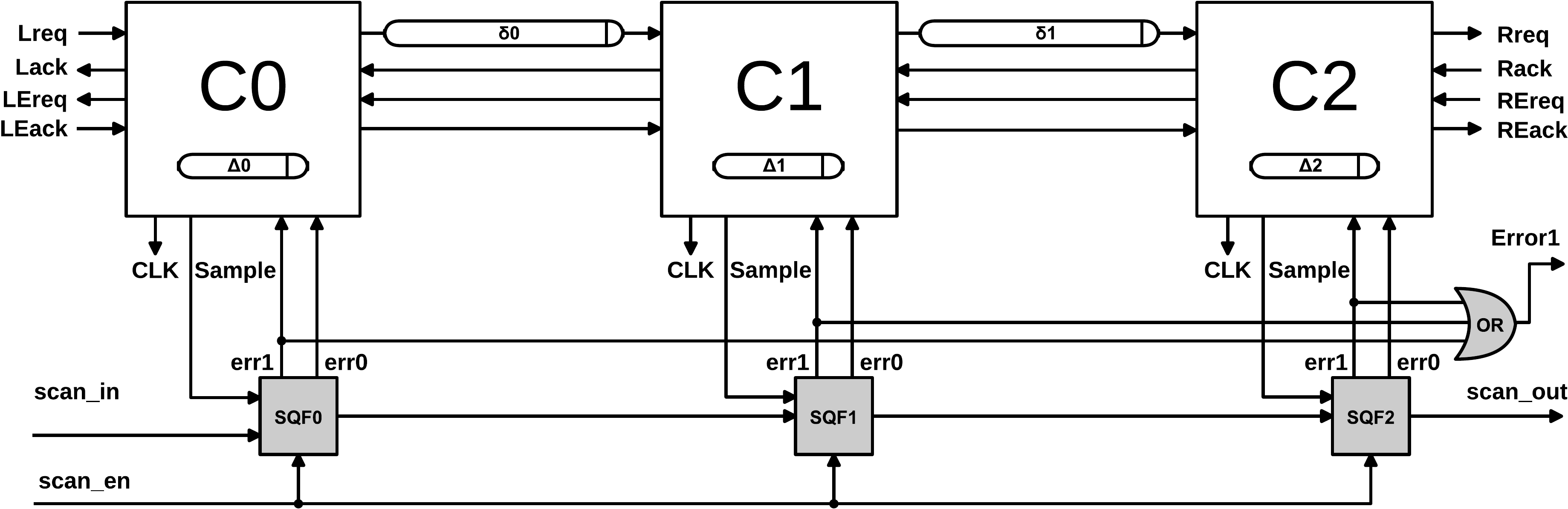}
    \caption{Delay lines test scenario with Blade design.}
    \label{fig:delay_test}
    \vspace{-1.2em}
\end{figure*}

%AMA acho essa frase importante p depois, nao aqui
%This test must be executed before delay testing of other structures, such as the combinational logic.

\section{Proposed Method}

The proposed method aims to use only primary inputs and outputs to take delay measurements, which include the handshake protocol pins (L/R channel and LE/RE channel) and an additional pin that serves as an observation point for the \textit{err1} signals, as illustrated in Fig. \ref{fig:delay_test}. Besides that, a scan chain to add controllability over the Q-Flops state is assumed. As previously discussed, a scannable data path is undesired due to high area overheads, but instead, the proposed method assumes the replacement of Q-Flops for Scan Q-Flops (SQF) \cite{JURA18}, which represents a smaller number of scan elements than a full scan. For instance, a 32 bit Blade \cite{BLADE15} stage requires 4 SQFs. The SQF is mandatory for the proposed test method since the Q-Flop acts as a metastability filter, and there is an unbounded time for metastability to be resolved inside the Q-Flop filter and settle the correct output error signal. With the SQF, this time uncertainty no longer exists, and correct timing assumptions can be made for the proposed test method.

\subsection{Test Architecture}

Fig. \ref{fig:delay_test} shows a simplified 3-stage pipeline example of the architecture. In this scenario, the left channels of C0 and the right channels of C2 are connected to primary inputs and outputs, and a single SQF representation is presented. Through the proposed scan chain \textit{err1} and \textit{err0} can be forced to a controlled state immediately after \textit{Sample} rises. Internal \textit{err1} signals are grouped with an OR gate, and its output is mapped to a primary output pin called \textit{Error1}. This design allows delay measurement of the \textit{err1} rising moment for the different stages. For instance, if SQF1 is configured to raise its \textit{err1} output signal, and all the others to rise their \textit{err0} signal, the rising edge observed at the primary output corresponds to the \textit{err1} connected to controller C1, assuming no fault at the OR gate.

%it is assumed that only \textit{ScanQflop1} (Fig.  \ref{fig:delay_test})

The proposed Blade controller \cite{BLADE15} implements the handshake protocol assuming two $\Delta$ delay lines and their consolidation into a single one is left for future works. For our test method, we assume that the $\Delta$ delay line is a single delay element.
%AMA bah, essa parte eh PURO ACHISMO. nada cientifico nisso. me parece que pode (ou ate, DEVE) ser  removida.
%FAK não entendi a parte do achismo!! eu afirmo que usando esse método NÂO tem como ter esses resultados se não for apenas uma linha de delay...se eu não disser que é apenas uma linha dou a entender que vai funcionar com o bleade original, e isso não é verdade... da mesma forma não da pra pegar e usar diretamente no SHARP...como dito abaixo
Nevertheless, the proposed test method could be extended for testing controllers with different internal delay lines such as Sharp \cite{SHARP17}. However, further analyses are required as the handshake protocol is different, and the time measurements need to be derived and measured from different pins and internal observation points. Accordingly, this is left for future work, along with an extension to consider forks and joins.

\subsection{Test Procedure}

Some important aspects of the controller and the implemented communication protocol must be described for a better understanding of the test procedure. The $\Delta$ delay line controls the high phase of \textit{CLK} and $\delta$ delay line is equivalent to the longest path of the combinational logic between the pipeline stages. %During the high phase of \textit{CLK} Blade's error detection logic evaluates if a timing violation has occurred and than \textit{Sample} rises to store the error signal at a Q-Flop and flags to to the controller if a timing violation was captured. The Q-Flop ovoid that metastability is propagated to the control logic. 
A request arriving at the primary input \textit{Lreq} is immediately forwarded from the left channel to the right channel connected to C0. This means that if no timing violations are flagged (only \textit{err0} rises), and the pipeline is empty, the time from \textit{Lreq} to \textit{Rreq} is equal to the sum of all $\delta$ delay lines propagation ($\delta0$, $\delta1$ and $\delta2$). Another thing to notice is that at the same time that \textit{Lreq} is flagged, internally the \textit{CLK} signal rises and the $\Delta$ delay line is activated. 
%AMA nao explicou a implicacao desse notice
%FAK Lreq implica na ativiação de CLK e delta... pra mim ta bem claro isso

%AMA paragrafo denso. nao tem como apresentar de uma forma mais simples?
%AMA gasta-se muito tempo justificando a escolha do err1 ao inves do err0. acho q nao precisa isso.
%FAK gasta-se pq é MUITO importante, o método não funcionaria se err0 fosse usado, o problema é explicar em detalhas o pq com o pouco espaço disponível
After the high phase of \textit{CLK} the \textit{Sample} signal rises, and \textit{err0} or \textit{err1} are immediately flagged, which is possible with the use of the SQF previously configured through the scan chain. In this case, it is possible to assume that the time from \textit{Lreq} to \textit{err1} or \textit{err0} is equal to the $\Delta$ delay line propagation, thus both can be used as an observation point to measure the $\Delta$ delay. However, \textit{err1} activates the protocol extension phase, which is required for measuring the $\delta$ delay, so \textit{err1} has been selected because it serves both tests. %Thus choosing \textit{err1} as the observation point will extend the delay, and the delay measurements will detect the additional time that corresponds to the $\Delta$ delay extension. 
 
One last thing to mention is that the \textit{REack} is the protocol signal that indicates that all delay lines ended their signal propagation, regardless of which error signal was flagged. Now it is possible to describe the test steps and the equations used to calculate the propagation delay of all delay lines in a circuit implemented with Blade. Between each step, the circuit goes through a complete reset.

\begin{itemize} 
    \item \textbf{Step 1:} The first delay measurement corresponds to the sum of all $\delta$ delays. All SQFs are configured to raise their \textit{err0} output signal and the communication is started at the left channel. Equation \ref{eq:total_deay} defines that the sum of all $\delta$ delays is the difference time between the first \textit{Lreq} transition ($T_{Lreq}$) and the first \textit{Rreq} transition ($T_{Rreq}$) that arrives at the end of the pipeline.
\end{itemize} 

\vspace{-0.8em}

\begin{equation} \label{eq:total_deay}
    T_{Sum} = T_{Rreq} - T_{Lreq}
\end{equation}

\begin{itemize}    
    \item \textbf{Step 2:} This step is repeated for each controller to take the delay measurement of the internal $\Delta$ delay line. For this test only the SQF connected to the target controller is configured to raise its \textit{err1} output signal. By forcing the delay extension, the pipeline propagation time is increased by the $\Delta$ delay. Equation \ref{eq:big_delta} is used to calculate the $\Delta$ delay of each controller. The difference time between the first \textit{Lreq} transition ($T_{Lreq}$) and the first \textit{REack} transition ($T_{REack}$) that arrives at the end of the pipeline, minus the sum of all $\delta$ delays
    %AMA nao seria melhor dar uma var para o termo 'the sum of all $\delta$ delays' ?
    %FAK já tinha feito assim, não sei pq mudei... vou colocar uma var
    calculated in the previous step, is equal to the $\Delta$ delay of the target controller $i$.
\end{itemize} 
%AMA qual eh o propostio do '=1' em 'err1_{i} = 1' ?
%FAK é dizer que só sinal de err1 do estagio i foi configurado pra subir... mas realmente não sei se isso aparece na equação... eu descrevo isso no texto... por hora vou remover

\vspace{-0.8em}

\begin{equation} \label{eq:big_delta}
    \Delta_{i} = T_{REack} - T_{Lreq} - T_{Sum} %\text{ , for } err1_{i} = 1  
\end{equation}

\begin{itemize}    
    \item \textbf{Step 3:} The last step is used for measuring the $\delta$ delay lines between controllers. This step is repeated for each $\delta$ delay line. For this test the SQF connected to the controller that follows the target delay line is configured to rise the \textit{err1} signal, and the transition observed in \textit{Error1} corresponds to this SQF. Equation \ref{eq:little_delta} is used to calculate the $\delta$ delay between each each stage. The target $\delta$ delay $i$ is given by the difference time between the first \textit{Lreq} transition ($T_{Lreq}$) and the first \textit{Error1} transition ($T_{Error1}$), minus the sum of $\delta$ delays previously calculated in this current step, minus the $\Delta$ delay of the controller that follows the target delay line ($i + 1$). This last subtraction accounts for the fact that the $i + 1$ \textit{err1} only rises after the high phase of \textit{CLK}, which is controlled by the $\Delta_{i+1}$.
\end{itemize}

\vspace{-0.8em}

\begin{equation} \label{eq:little_delta}
    \delta_{i} = T_{Error1} - T_{Lreq} - (\sum_{k=0}^{i-1} \delta_{k}) - \Delta_{i+1} %\text{ , for } err1_{i+1} = 1
\end{equation}

\section{Preliminary Results}

An RTL description of the pipeline presented in Fig. \ref{fig:delay_test} was used for validation of the proposed test procedure and equations, and all the delays were correctly extracted. For these tests only the delays lines had propagation delay values.

The estimated area of the system described in Fig. \ref{fig:delay_test} is $Area = N * (A_{control} + A_{Qflop}) + A_{NinOR}$, where it represents the area of the controller, the SQF cell, and the N-input OR gate, respectively. N is the number of pipeline stages. The delay line area is excluded since it is dependent on the design. Assuming a fabrication technology of \SI{28}{\nano\meter}, the original Q-Flop has an area of about \SI{7}{\micro\meter\squared} with 28 transistors, while the SQF cell has an area of \SI{10}{\micro\meter\squared} with 40  transistors. The controller presents an area of \SI{27}{\micro\meter\squared}, and the OR presents an area of \SI{2.6}{\micro\meter\squared}. Thus, for the scenario presented in Fig.~\ref{fig:delay_test}, the total area is \SI{113.6}{\micro\meter\squared}, while the area without the test (i.e. using the Q-Flop cell) is \SI{102}{\micro\meter\squared}, representing an overhead of 11.37\%.

\section{Conclusions}

This paper proposed a method for testing the delay lines of Blade. The method correctly extracts the Blade's delay lines values and presents a low area overhead of 11.37\%. The next step is to apply the proposed method to a mapped netlist in order to account for additional propagation delays and the related error introduced in the time measurements.

\bibliographystyle{IEEEtran}
\bibliography{async19}

\end{document}